\documentclass[12pt]{article}

\usepackage{authblk} 
\usepackage{graphicx}
\usepackage{epsfig}
\usepackage{dcolumn}
\usepackage{bm}
\usepackage{amsmath}
\usepackage{amssymb}

\newcommand{\eqnref}[1]{Eq.~(\ref{eqn:#1})}
\newcommand{\eqnsref}[2]{Eqs.~(\ref{eqn:#1}) and (\ref{eqn:#2})}
\newcommand{\secref}[1]{Sec.~\ref{sec:#1}}

\newcommand{\figref}[1]{Fig.~\ref{fig:#1}}

\newcommand{\drawsquare}[2]{\hbox{%
\rule{#2pt}{#1pt}\hskip-#2pt
\rule{#1pt}{#2pt}\hskip-#1pt
\rule[#1pt]{#1pt}{#2pt}}\rule[#1pt]{#2pt}{#2pt}\hskip-#2pt
\rule{#2pt}{#1pt}}

\newcommand{\Yasymm}{\drawsquare{7}{0.6}\hskip-7.6pt%
\raisebox{7pt}{\drawsquare{7}{0.6}}}

\def\oraf{O'Raifeartaigh}
\def\Luv{\Lambda_{UV}}
\def\lsim{\lower0.5ex\hbox{$\:\buildrel <\over\sim\:$}}
\begin{document}

\begin{titlepage}
\begin{flushright}
ICRR-REPORT-586-2011-3 \\
UCI-TR-2011-04
\end{flushright}

\vskip.5cm
\begin{center}
{\huge \bf
R-symmetry Matching In SUSY Breaking Models
}

\vskip.1cm
\end{center}
\vskip0.2cm

\begin{center}
{\large \bf Jessica Goodman$^{a}$,
Masahiro Ibe$^{b}$,\\
Yuri Shirman$^a$, {\rm and}
Felix Yu$^{a}$}
\end{center}
\vskip 8pt

\begin{center}
$^{a}$ {\it
Department of Physics, University of California, Irvine, CA
92697.} \\

$^{b}$ {\it Institute for Cosmic Ray Research, University of Tokyo,
    Chiba 277-8582, Japan} \\

\vspace*{0.3cm} {\tt \footnotesize j.goodman@uci.edu,
  ibe@icrr.u-tokyo.ac.jp,\\ yshirman@uci.edu, felixy@uci.edu}
\end{center}

\vspace*{0.2cm}

\vglue 0.3truecm

\begin{abstract}
\vskip 3pt \noindent Low energy descriptions of metastable
supersymmetry breaking models often possess an accidental
R-symmetry. Viable phenomenological applications of this class of
models require R-symmetry to be broken in the ground state. This can
be achieved in \oraf-like models where some of the chiral superfields
carry negative R-charges. In this paper we consider UV completions of
this class of models and formulate necessary conditions that they must
satisfy. We show that the R-symmetry of the IR description can be
traced to an anomalous or anomaly-free R-symmetry of the UV theory and
discuss several representative examples.
\end{abstract}


\end{titlepage}
\newpage



\section{Introduction}
\label{sec:intro}
Supersymmetry has been extensively explored as one of the most
plausible extensions of the Standard Model at the TeV scale. One of
its most attractive features is the potential to solve the gauge
hierarchy problem.  Models with softly broken supersymmetry guarantee
that the electroweak scale is radiatively stable, thus providing a
solution for the technical naturalness problem. In addition, if
supersymmetry is broken dynamically, such breaking is necessarily
non-perturbative and the SUSY breaking scale is naturally small
compared to the fundamental scale of the theory (such as the Grand
Unification scale or the Planck scale).  Understanding of the
dynamical supersymmetry breaking is especially important in scenarios
with gauge mediated supersymmetry breaking (GMSB).  Indeed, in GMSB
models, supergravity contributions to the parameters of the low energy
Lagrangian are negligible and studies of the full theory, including
both the Standard Model and SUSY breaking sectors, may be under full
theoretical control.  Moreover, in models with extremely low SUSY
breaking scales (of order tens or hundreds of TeV), many new particles
or interactions may be experimentally accessible.

However, it is difficult to find DSB models with phenomenologically
desirable features. Moreover, once the DSB sector is coupled to the
Standard Model, the SUSY breaking vacuum generically survives only as
a local minimum of the potential. This is acceptable as long as the
lifetime of the metastable vacuum is sufficiently long. The model
building prospects improved significantly when Intriligator, Seiberg,
and Shih (ISS)~\cite{Intriligator:2006dd} proposed to embrace
metastability as a fundamental feature of the models. They showed that
metastable, long-lived (and often calculable) non-supersymmetric vacua
are generic in SUSY gauge theories. ISS models usually possess an
accidental R-symmetry which is unbroken in the metastable vacuum.
This poses a significant obstacle to constructing viable extensions of
the Standard Model since an unbroken R-symmetry forbids gaugino
masses.  Several implementations of direct gauge mediation based on
ISS models proposed in the
literature~\cite{Dine:2006xt,Csaki:2006wi,Kitano:2006xg} circumvent
this difficulty by modifying the underlying theory so that R-symmetry
is broken either explicitly or spontaneously.  Nevertheless, it was
argued in~\cite{Giveon:2009yu} that even then gaugino masses are
numerically suppressed compared to sfermion masses.  The phenomenology
of low scale GMSB models with spontaneous and explicit R-symmetry
breaking as well as the ability to discriminate these two classes was
investigated in~\cite{Abel:2007nr}.

Motivated by these problems, Shih studied conditions for spontaneous
R-symmetry breaking in O'Raifeartaigh-like models~\cite{Shih:2007av}.
He found that the Coleman-Weinberg potential may result in
simultaneous supersymmetry and R-symmetry breaking if the model
contains chiral superfields with R-charges other than $0$ and
$2$. This requirement implies that models must contain at least one
chiral superfield with a negative R-charge.  Although the
O'Raifeartaigh description is often sufficient for phenomenological
purposes, it is desirable to understand R-symmetry breaking dynamics
in terms of a complete UV description\footnote{See~\cite{Abel:2007jx}
  for examples of UV completions.}.  The requirement of negative
R-charges suggests that searching for a UV description may be
tricky. Indeed, the presence of negative R-charges allows one to write
superpotential terms with negative exponents of the superfields.  Such
terms must be forbidden by the symmetries of the microscopic physics
--- otherwise they would necessarily be generated dynamically and
destabilize the SUSY breaking minimum.

In this paper we will study general requirements for UV completions of
\oraf-like models with perturbative R-symmetry breaking.  We will show
that in viable models the low energy R-symmetry arises as a linear
combination of a (possibly anomalous) R-symmetry of the UV physics and
an anomaly-free global symmetry. In the case of an anomalous
R-symmetry, R-charges of all physical fields in the microscopic
description will be anomalous. When the R-symmetry is anomaly-free,
R-charges of the low energy fields arise due to contributions of the
non-R global symmetry. In this case dangerous operators do not appear
since the non-R global symmetry is respected by all the
non-perturbative dynamics.

\section{General properties of UV completions}
\label{sec:gen_props}
We are interested in models of metastable SUSY breaking where some of
the low energy degrees of freedom are composites of the microscopic
physics~\footnote{We explicitly exclude from consideration
  retrofitting models~\cite{Dine:2006gm} where the role of the
  non-perturbative dynamics is restricted to generation of mass
  parameters in the superpotential. On the other hand, our results can
  be easily generalized to models where IR and UV degrees of freedom
  are related by a duality transformation.}.  There exist several
possibilities for the origin of the R-symmetry of the low energy
physics:
\begin{itemize}
\item An R-symmetry of the low energy description arises from the
  non-anomalous R-symmetry of the microscopic physics. In this case,
  the R-symmetry is a symmetry of the Lagrangian at all scales.  Many
  classic models of dynamical supersymmetry breaking belong to this
  class. They possess R-symmetries under which some superfields carry
  negative charges. Non-perturbative dynamics generates superpotential
  terms with negative powers of the superfields. Both SUSY and
  R-symmetry are broken due to the interplay between tree level and
  non-perturbative interactions while perturbative corrections are
  small. In Sec.~\ref{subsec:modelA}, we will show that this class of
  models also contains theories where negative exponents of
  superfields do not appear in the dynamical superpotential and
  R-symmetry is broken by perturbative dynamics.
\item An R-symmetry of the IR physics corresponds to an anomalous
  R-symmetry of the ultraviolet description.  Since the R-symmetry is
  anomalous, it will necessarily be broken by non-perturbative
  dynamics. As we will show, the existence of the negative R-charges
  in the low energy description implies that the microscopic theory
  contains at least some elementary chiral superfields with negative
  R-charges. Thus, in general, it is possible that the superpotential
  terms containing fields with negative exponents will be generated by
  the non-perturbative dynamics. Such terms can never be negligible
  near the origin of the moduli space. In models with classical flat
  directions they tend to destabilize the SUSY breaking vacuum of the
  O'Raifeartaigh-like model. However, we will construct models where
  an anomalous R-symmetry is given by a linear combination of
  anomalous and anomaly-free symmetries ({\it e.g.} baryon
  number). While certain composites carry negative R-charge, they only
  show up in the dynamical superpotential in combination with fields
  carrying sufficiently large positive R-charge so that the dynamical
  superpotential does not contain superfields with negative
  exponents. As a result, the SUSY and R-symmetry breaking minimum of
  an O'Raifeartaigh model will survive as a local minimum of the UV
  completion. We will discuss the correspondence between an accidental
  R-symmetry of the effective low energy description and an anomalous
  R-symmetry of the microscopic theory in Secs.~\ref{subsec:modelB}
  and~\ref{subsec:modelC}.
\end{itemize}

\section{Anomalous R-symmetries and non-perturbative superpotentials in SQCD}
\label{sec:SQCD}
Let us briefly review exact results from supersymmetric QCD. For our
purposes it is convenient to follow the presentation
of~\cite{Peskin:1997qi}.  Consider an $SU(N)$ gauge theory with $F$
flavors in the fundamental representation. The quantum numbers of the
fields under gauge and global symmetries are:

\begin{equation}
\begin{array}{c|c| c c c c c}
         & SU(N)_{\text{gauge}} & SU(F)_L    & SU(F)_R 
         & U(1)_B & U(1)_A & U(1)_{R} 
\\
\hline
Q      & \Box             & \Box       & \mathbf{1} 
         & 1      & 1     & 0 \\
\bar Q & \overline{\Box}  & \mathbf{1} & \overline{\Box} 
         & -1     & 1     & 0 \\
\hline
\end{array}
\label{eqn:SQCDreps}  
\end{equation}

Both the $U(1)_A$ and the $U(1)_R$ symmetries are anomalous.  If we
perform the corresponding symmetry transformation parametrized by an
angle $\alpha$, each fermion transforming under representation $r$ and
carrying charge $q_r$ under the anomalous symmetries will contribute a
factor of $n_r\alpha F\tilde F$ to a shift in the Lagrangian, where
$n_r$ is an anomaly coefficient given by
\begin{equation}
n_r =  2q_rC(r)= \left\{ \begin{array}{ll} 
q_r  & r = \Box \text{ or } \overline{\Box} \\ 
2Nq_r & r = \text{ adjoint} \ . \\
\end{array} \right.
\end{equation}
This shift can be absorbed into a redefinition of the $\theta$-angle,
$\theta\rightarrow \theta-\sum_{r}n_r\alpha$, thus formally restoring
the symmetry.  In SUSY gauge theories, the gauge coupling and the
$\theta$-angle combine into a holomorphic background superfield $\tau$
given by
\begin{equation}
\tau = \frac{\theta}{2\pi}+\frac{4\pi i}{g^2} \ .
\end{equation}
Thus quantum physics remains formally invariant under anomalous
symmetries if the gauge function $\tau$ transforms non-linearly:
\begin{equation}
\tau \rightarrow \tau-\frac{\sum_rn_r\alpha}{2\pi}\,.
\end{equation}
For example, in the case of an anomalous R-symmetry, as defined
in~\eqnref{SQCDreps}, the fermions in the quark supermultiplets carry
R-charge $-1$, while the R-charge of gauginos is $1$. Thus $\sum
n_r=2(N-F)$.  On the other hand, the renormalization group evolution
of the gauge coupling allows us to associate $\tau$ with the dynamical
scale of the theory using
\begin{equation}
\Lambda^{b_0} = M^{b_0} e^{2\pi i \tau(M)} \ ,
\end{equation}
where $b_0=3N-F$ is a one loop beta function coefficient of SUSY QCD
and $\tau(M)$ is a running coupling evaluated at the scale $M$. We see
that non-linear transformations of $\tau$ under the R-symmetry
correspond to linear transformations of $\Lambda^{b_0}$ with R-charge
$2(N -F)$. If we now specialize to models with $F < N$, we can easily
see that the function
\begin{equation}
\label{eqn:ADSW}
W = \left(\frac{\Lambda^{b_0}}{\det Q\bar Q}\right)^{\frac{1}{N-F}}
\end{equation}
has an R-charge of 2. Similar arguments lead to the conclusion that
$W$ is invariant under $U(1)_A$ if $\Lambda^{b_0}$ carries charge $2F$
under $U(1)_A$. Indeed, the superpotential of~\eqnref{ADSW} is the
celebrated non-perturbative Affleck-Dine-Seiberg (ADS) superpotential
(in the case $F=N-1$, the overall coefficient of this term can be
evaluated by an explicit instanton calculation).  Applying this
formalism to \oraf-like models, we will be able to relate the
R-symmetry of the IR description to anomalous or anomaly-free
symmetries of the UV physics.

\section{Metastable dynamical SUSY breaking with spontaneously broken 
R-symmetry}
\label{sec:models}
It is well known that O'Raifeartaigh models possess pseudoflat
directions in the field space which can be parameterized by vacuum
expecation values (vevs) of moduli fields with R-charge 2. Thus, at a
generic point on the pseudomoduli space, R-symmetry is spontaneously
broken. Perturbative corrections lift the pseudoflat direction and
generically stabilize pseudomoduli at the origin so that R-symmetry is
unbroken in the ground state. It was shown in~\cite{Shih:2007av} that
O'Raifeartaigh models can be generalized to a class of models where
the Coleman-Weinberg potential results in the existence of a local
minimum of the potential with spontaneously broken R-symmetry (albeit
the SUSY breaking minimum is only a local one).  Here we will discuss
several UV completions of models in this class and will show how the
R-symmetry of the effective description arises from the symmetries of
the microscopic theory.

For simplicity we will study $SU(N)$ SQCD with $F$ flavors.  As we
have seen in~\secref{SQCD}, the maximal global symmetry is $SU(F)_L
\times SU(F)_R \times U(1)_B \times U(1)_A \times U(1)_R$, where
$U(1)_A$ and $U(1)_R$ are anomalous.  If we require that
non-perturbative dynamics does not generate superpotential terms that
are singular at the origin of the moduli space, we must have $F > N -
1$. On the other hand, to simplify the discussion we will assume that
there are no light gauge fields in the IR description which implies $F
< N + 2$.

In Sec.~\ref{subsec:modelA} we study a model with $F = N$ and explain
how an accidental R-symmetry of the IR physics can arise from an
anomaly-free R-symmetry of the UV theory. In Secs.~\ref{subsec:modelB}
and~\ref{subsec:modelC}, we consider the case with $F=N+1$ and
identify the R-symmetry of the low energy physics with an anomalous
R-symmetry of the UV completion.

\subsection{A model with a non-anomalous R-symmetry}
\label{subsec:modelA}
The simplest SUSY and R-symmetry breaking model, introduced
in~\cite{Shih:2007av}, has the superpotential
\begin{equation}
\label{eqn:modelAlow}
W = \lambda X (\mu^2 -\phi_1\phi_2)+m_1 \phi_1 \phi_3+
\frac{m_2}{2}\phi_2^2 \ .
\end{equation}
The model possesses an R-symmetry with the charges of the chiral
superfields given by
\begin{equation}
\label{eqn:modelAlowfieldid}
R(X) = 2, \quad R(\phi_1) = -1, \quad R(\phi_2) = 1, \quad 
R(\phi_3) = 3 \ .
\end{equation}

Our UV completion will be based on a perturbation of the ITIY
model~\cite{Intriligator:1996pu,Izawa:1996pk} with an $SU(2)$ gauge
group, 4 doublet chiral superfields $Q_i$ and 6 gauge singlet fields
$S_{ij}$ transforming under gauge and global symmetries as
\begin{equation}
\begin{array}{c|c| c c  }
& SU(2)_{\text{gauge}} & SU(4) & U(1)_{R} \\ [0.5ex] 
\hline&&& \\                
Q            & \Box & {\Box} & 0\\   
S            & 1 & \Yasymm
&2\\ 
\hline&&&
\end{array}
\end{equation}
The classical superpotential is chosen to be
\begin{equation}
\label{eqn:modelAUV}
W = \sum\limits_{i, j = 1, \ i < j}^4 \lambda_{ij} S_{ij} Q_i Q_j 
  + \frac{(Q_3Q_4)^2}{\Lambda_{UV}} + \frac{m_S}{2} S_{34}^2 \ .
\end{equation}

The second two terms in~\eqnref{modelAUV} break the maximal
anomaly-free global symmetry down to $SO(4) \times U(1)^\prime_{R}$,
where the unbroken anomaly-free R$^\prime$-symmetry is a linear
combination of the original R-symmetry and the $U(1)_F$ subgroup of
$SU(4)$ generated by $T = \text{diag}(-1, -1, 1, 1)$.  The
R$^\prime$-symmetry charges are given by
\begin{equation}
\begin{split}
& R^\prime(Q_1) = R^\prime(Q_2) = -\frac{1}{2} \ , \quad R^\prime(Q_3) 
  = R^\prime(Q_4) = \frac{1}{2} \ , \\ 
& R^\prime(S_{12}) = 3 \ , \quad R^\prime(S_{34}) = 1 \ , \quad 
  R^\prime(S_1) = R^\prime(S_2) = R^\prime(S_3) = R^\prime(S_4) = 2 \ .
\end{split}
\end{equation}

One of the important features of the ITIY model is the absence of
classical flat directions involving quark superfields. A quick
analysis of~\eqnref{modelAUV} shows that for generic choices of the
tree-level parameters this remains true in the presence of our
perturbation. Classical flat directions are reintroduced when
$\lambda_{34}^2 / (2 m_S) - 1 / \Lambda_{UV} = 0$, leading to
restoration of supersymmetry.

Let us show the correspondence between the models defined
by~\eqnsref{modelAlow}{modelAUV}. The full superpotential of the
perturbed ITIY model is
\begin{equation}
W = \chi (\text{ Pf } M - \Lambda^4) + \sum_{ij} \lambda_{ij} S_{ij}
M_{ij} + c \frac{M_{34}^2}{\Lambda_{UV}} + \frac{m_S}{2} S_{34}^2 \ ,
\end{equation}
where $\chi$ denotes the Lagrange multiplier which represents the
quantum deformed moduli constraint.  Our perturbation singles out two
mesons, $M_{12}=(Q_1 Q_2)$ and $M_{34}=(Q_3Q_4)$, and their associated
singlets. We will use the unbroken $SO(4)$ symmetry to denote the
remaining mesons and singlets as $M_a$ and $S_a$, $a = 1, \ldots 4$,
respectively. Using the quantum constraint, we can integrate out one
of the mesons, say $M_1$,
\begin{equation}
M_1 = \left( \Lambda^4 - \sum\limits_{a=2}^4 M_a^2 - 2 M_{12} M_{34}
\right)^{1/2} 
\simeq \Lambda^2 - \sum\limits_{a=2}^4
\frac{M_a^2}{2 \Lambda^2} - \frac{ M_{12} M_{34}}{\Lambda^2} + \ldots
\end{equation}
where the dots represent higher order terms in the expansion.  The
superpotential becomes
\begin{equation}
\label{eqn:modelAlow2}
\begin{array}{ccl}
W &=& \lambda_1 S_1 \left(\Lambda^2 - \sum\limits_a \frac{M_a^2}{2
  \Lambda^2} - \frac{M_{12} M_{34}}{\Lambda^2} \right) \\
  &+& \sum\limits_a \lambda_a S_a M_a + \lambda_{12} S_{12} M_{12} 
+ \lambda_{34} S_{34} M_{34} + c \frac{ M_{34}^2}{\Lambda_{UV}} +
\frac{m_S}{2}S_{34}^2 \ .
\end{array}
\end{equation}
Once we integrate out the massive fields, $M_a$, $S_a$, and $S_{34}$,
the correspondence between UV and IR descriptions becomes obvious:
\begin{equation}
\label{eqn:modelAfieldid}
X \sim S_1, \quad \phi_1 \sim M_{12}/\Lambda, \quad \phi_2 \sim
M_{34}/\Lambda, \quad \phi_3 \sim S_{12} \ .
\end{equation}
Thus the pseudomoduli spaces of the two models are identical. 

\begin{figure}[htb]
\caption{[color online].  Relative Coleman-Weinberg potential as a
  function of pseudomodulus $S_1$ for the model in~\eqnref{modelAlow2}
  where $\Lambda = 1$, $\Lambda_{UV} = 10$, $\lambda_1 = 0.02$,
  $\lambda_a = 1$, $\lambda_{12} = 0.03$, $\lambda_{34} = 0.03$, $m_S
  = 1$, and $c$ varied from 0.2 (blue, dotted, top) to 0.6 (purple,
  solid, middle) to 1.0 (green, dashed, bottom).}
\label{fig:modelACW}
\begin{center}
\includegraphics[width = 0.5\textwidth]{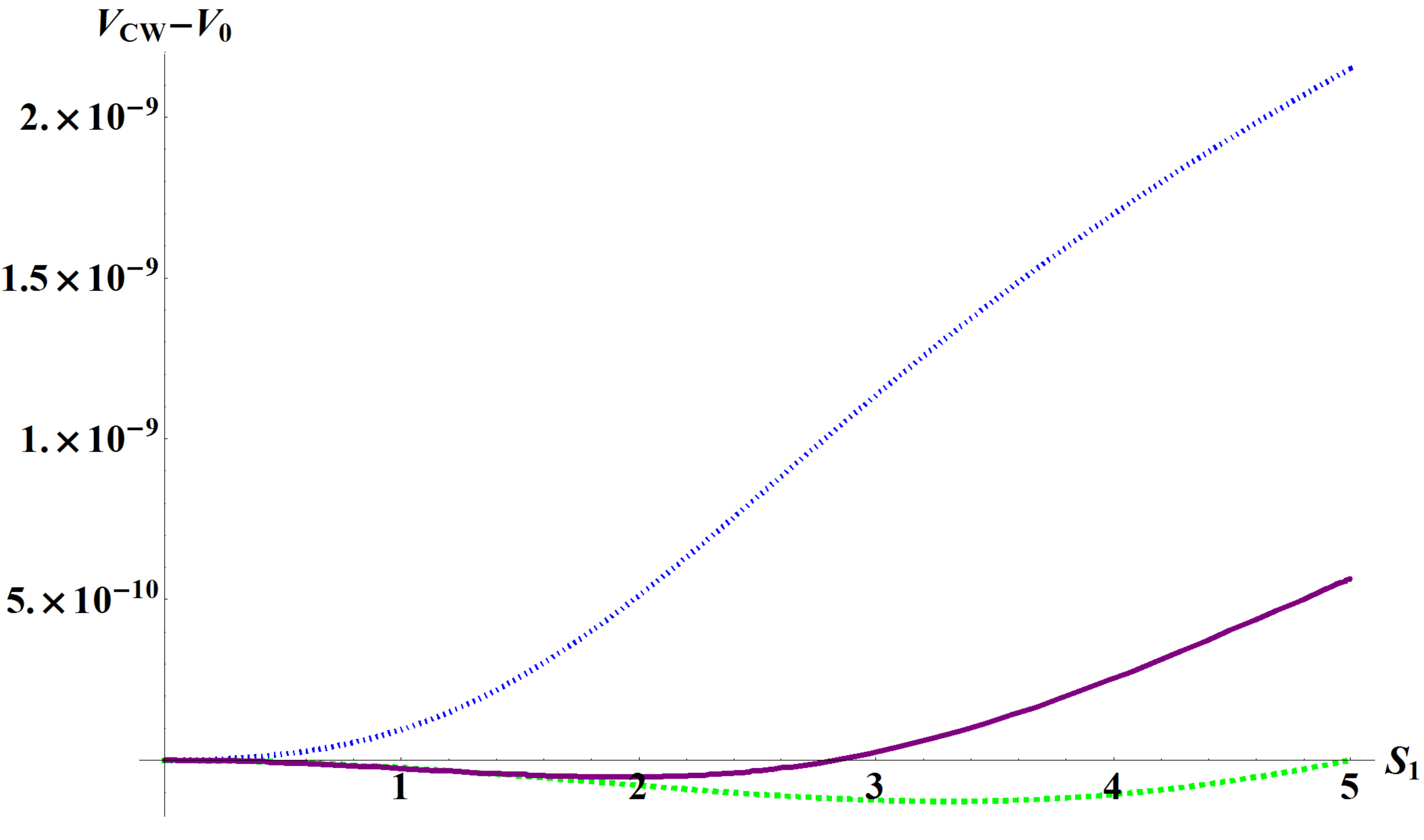}
\end{center}
\end{figure}

It is important to note that relation~\eqnref{modelAfieldid}
constrains the coupling constants and masses of fields in the low
energy effective description according to
\begin{equation}
\mu \sim \Lambda, \quad 
\lambda \sim \lambda_1, \quad 
m_1 \sim \lambda_{12} \Lambda, \quad 
m_2 \sim \left( \frac{1}{\Lambda_{UV}} - \frac{\lambda_{34}^2}{2 m_S} 
\right) \Lambda^2 \ .
\end{equation}
In the model from~\eqnref{modelAlow}, SUSY is restored when $m_2$ is
massless. From the point of view of the microscopic description, this
happens precisely when UV parameters are chosen so that the classical
flat direction is reintroduced. On the other hand, for a generic
choice of the parameters, the local SUSY breaking minimum exists and
for a range of parameters, R-symmetry is broken in this vacuum
(see~\figref{modelACW}).

The model we discussed here clearly exemplifies the importance of the
UV completion, which possesses a larger set of anomaly-free global
symmetries. Indeed, while superpotential terms with negative
superfield exponents are allowed by all of the symmetries of the model
of~\eqnref{modelAlow}, such terms are forbidden by additional
symmetries present in the UV completion of~\eqnref{modelAUV}. In
particular, the R-symmetry of the low energy physics is a linear
combination of the $U(1)_R$ and vector-like $U(1)_F$ symmetries. In
fact, it is $U(1)_F$ which is responsible for the appearance of
composites with negative R-charges. At the same time, dynamical terms
in the superpotential must be invariant under all anomaly-free global
symmetries, thus preventing the appearance of the terms with negative
exponents of the superfields.

\subsection{A model with an anomalous R-symmetry}
\label{subsec:modelB}
We have seen in the previous section that global symmetries of the UV
physics play an important role in understanding of the IR dynamics of
\oraf-like models. Therefore, we will consider a generalization of the
model from~\eqnref{modelAlow} in the form
\begin{equation}
\label{eqn:modelB}
W = \lambda \phi_i X^{ij} \tilde{\phi}_j - \mu^2 \phi_1 
+ \frac{1}{2} m \text{ Tr } X^2 + n \tilde{\phi}_i S^i \ ,
\end{equation}
where $i,j=1, \ldots F,$ are flavor indices, $\mu$, $m$, and $n$ are
mass parameters, and $\lambda$ is a coupling constant. The model
possesses a large global symmetry, including an R-symmetry under which
chiral superfields carry the following charges~\footnote{Due to the
  presence of the large global symmetry, this definition of R-charges
  is not unique for some fields.}:
\begin{equation}
\label{eqn:modelBR}
R_\phi=2, \quad R_{\tilde{\phi}} = -1, \quad R_X=1, \quad R_S=3 \ .
\end{equation}

For a UV completion, we will consider $SU(N)$ theory with $F=N+1$
flavors and map $\phi_i$, $\tilde{\phi}_i$ and $X_{ij}$ to baryons,
anti-baryons and mesons of the microscopic description, respectively,
while $S_i$ remain elementary:
\begin{equation}
\label{eqn:fieldmap}
\phi_i \sim B_i \ , \quad \tilde{\phi}_i \sim \overline{B}_i \ , \quad
X_{ij} \sim M_{ij} \ .
\end{equation}

In the absence of the superpotential, the global symmetry is $SU(F)_L
\times SU(F)_R \times U(1)_B \times U(1)_A \times U(1)_R$.  Charges of
the matter fields, gauge invariant composites, and the dynamical scale
$\Lambda$ are given by

\begin{equation}
\begin{array}{c|c| c c c c c}
& SU(N)_{\text{gauge}} & SU(N+1)_L & SU(N+1)_R & U(1)_B & U(1)_A 
& U(1)_{R} \\ [0.5ex] 
\hline&&&&&& \\                
Q            & \Box & {\Box} &1&\frac{1}{N} &\frac{1}{N}& 0\\   
\bar Q         & \overline{\Box} & 1 &\overline{\Box} 
& -\frac{1}{N}&\frac{1}{N}&0 \\
S            &1&1&\overline{\Box}&1&-1&2\\ 
\hline&&&&&&\\
\Lambda^{2N-1}&&&&&\frac{2(N+1)}{N}&-2\\
\hline&&&&&&\\
B=Q^{N} & 1 & \overline{\Box} &1&1&1&0   \\ 
\overline B=\overline Q^{N} & 1 &1& \Box &-1&1&0   \\ 
M=Q\bar Q&1&\Box&\overline{\Box}&0&\frac{2}{N}&0\\
\hline     

\label{eqn:modelBreps}  
\end{array}
\end{equation}

To analyze the model from the microscopic point of view, we need to
choose a tree level superpotential that matches~\eqnref{modelB} as
closely as possible.  It is easy to check that the
superpotential~\eqnref{modelB} does not possess any R-symmetry when
written in terms of the elementary fields. An R-symmetry may appear if
the superpotential depends on the dynamical scale $\Lambda$ which
transforms under anomalous symmetries. Therefore, at least some terms
in~\eqnref{modelB} must be generated dynamically. Indeed, it is well
known~\cite{Seiberg:1994bz, Seiberg:1994pq} that $\phi X \tilde{\phi}
\sim B M \bar{B}/\Lambda^{2N-1}$ is generated non-perturbatively.  Let
us then restrict our attention to the remaining three terms
in~\eqnref{modelB}. If we require that these terms correspond to the
tree level superpotential of the microscopic description, we find an
anomalous R-symmetry given by
\begin{equation}
\label{eqn:RpmodelB}
U(1)^\prime_{R} = U(1)_R + \frac{N}{2} U(1)_A + (2 - \frac{N}{2})
U(1)_B \ .
\end{equation}

The full dynamical superpotential is
\begin{equation}
\label{eqn:uvB}
W = \lambda \frac{B_i M^{ij} \bar{B}_j - \det M}{\Lambda^{2N-1}} + 
c_B\frac{B_1}{\Luv^{N-3}} + c_M\frac{ \text{ Tr } M^2}{\Luv} 
+ c_{\bar{B}} \frac{\bar{B}_i S^i}{\Luv^{N-2}} \ ,
\end{equation}
where the dimensionless coefficients $c_B$, $c_M$, and $c_{\bar{B}}$
are of $\mathcal{O}(1)$ in the absence of fine-tuning (we will see
shortly that in this model one must choose $c_B\ll 1$). This
superpotential is invariant under~\eqnref{RpmodelB} once the
transformation properties of $\Lambda$ are taken into account.  An
additional term, $\det M$, appearing in~\eqnref{uvB} remains
irrelevant in the IR and decouples from low energy physics.

Since the existence of the R-symmetry in the UV description required
an addition of the spurion $\Lambda^{2N-1}$, the matching of R-charges
is not completely trivial. Charges of the tree-level terms $B_1$ and
$\text{Tr } M^2$ must match directly between the UV and IR. Comparing
the non-perturbative term to its counterpart in the IR superpotential,
we see that the charge of $\bar B/\Lambda^{2N-1}$ matches the charge
of $\tilde \phi$. This, in turn, determines the matching between UV
and IR charges of the gauge singlet fields. The full set of relations
between the R-charges is
\begin{equation}
\label{eqn:Rmatch}
R_\phi = R_B \ , \quad R_X = R_M \ , 
R_{\tilde{\phi}} = R_{\bar{B}} - R_{\Lambda^{2N-1}} \ , \quad 
R_{S_{IR}} = R_{S_{UV}} + R_{\Lambda^{2N-1}} \ .
\end{equation}
It is important to note that in our construction all the superfields
of the microscopic theory have positive R-charges.  Negative R-charges
of the low energy effective description are due to the contribution of
the anomaly through the spurion $\Lambda^{2N-1}$. This guarantees that
all terms generated by non-perturbative dynamics are regular at the
origin of the moduli space.

\begin{figure}[tbh]
\caption{[color online]. Relative Coleman-Weinberg potential as a
  function of pseudomodulus $B_1$ for~\eqnref{uvB} with $\Lambda = 1$,
  $\Lambda_{UV} = 10$, $N = 4$, $\lambda = 1$, $c_B = 0.1$, $c_M =
  4.0$, and $c_{\bar{B}}$ varied from 1.5 (green, dashed, bottom) to 2.5
  (purple, solid, middle) to 3.5 (blue, dotted, top).}
\label{fig:modelBCW}
\begin{center}
\includegraphics[width = 0.5\textwidth]{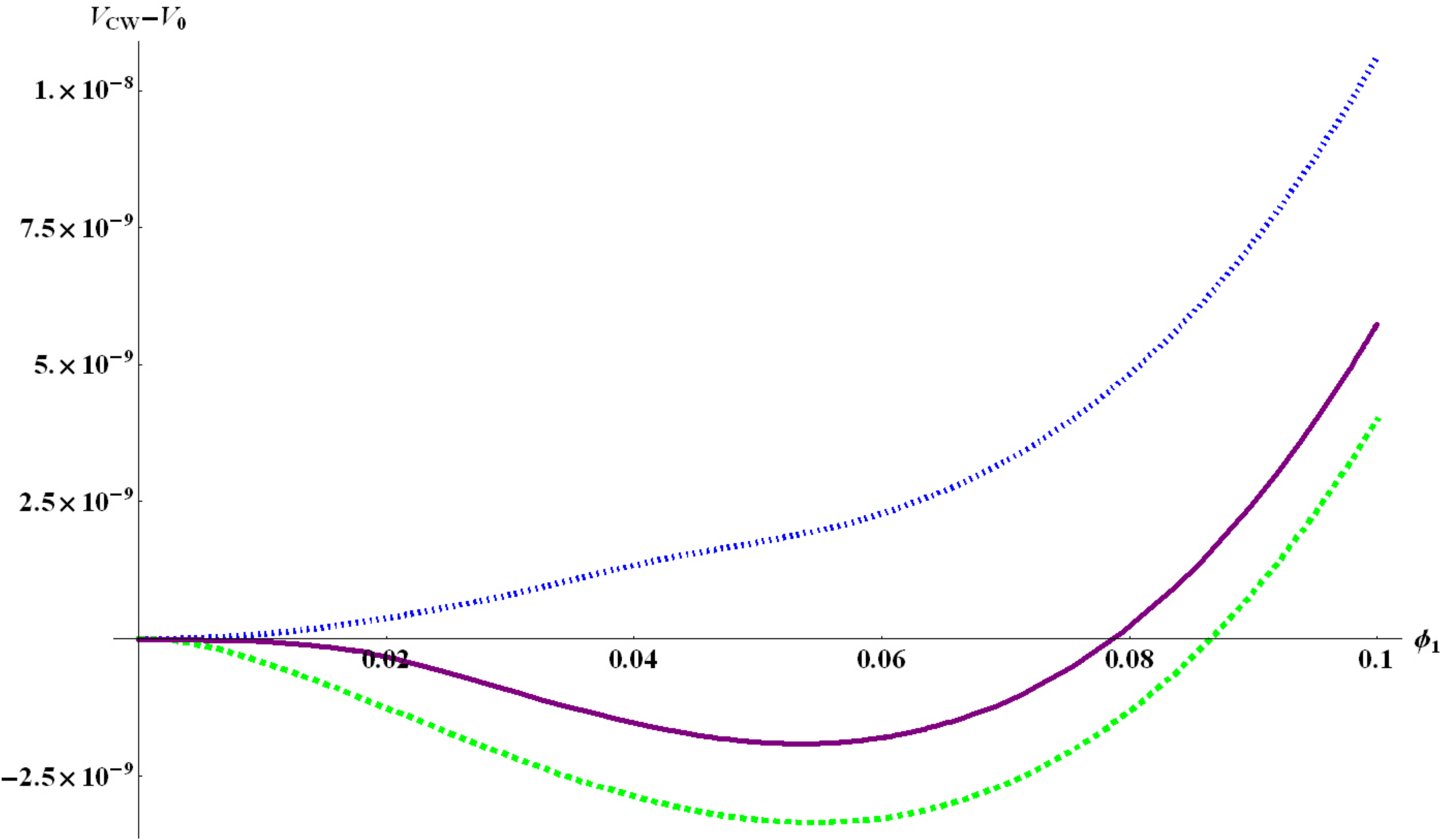}
\end{center}
\end{figure}

To study the Coleman-Weinberg potential, we first neglect the
non-renormalizable term $\det{M}$ in~\eqnref{uvB} and thus restrict
our attention to the superpotential~\eqnref{modelB}.  We also note
that the parameters of the low energy model are related to those of
the microscopic description according to
\begin{equation}
\mu^2 = c_B\left( \frac{\Lambda}{\Luv} \right)^{N-3} \Lambda^2 \ , \quad
\frac{m}{2} = c_M \left( \frac{\Lambda}{\Luv} \right) \Lambda \ , \quad
n = c_{\bar{B}} 
\left( \frac{\Lambda}{\Luv} \right)^{N-2} \Lambda \ ,
\end{equation}
and $\lambda \sim \mathcal{O}(1)$.  Otherwise, our analysis closely
follows that of~\cite{Shih:2007av} and arrives at the same
conclusions.  It is easy to see that at tree level the model possesses
a flat direction in the field space along which the energy is
non-vanishing, with scalar potential $V_{min} = \mu^4$. This direction
is parametrized by
\begin{eqnarray}
\label{eqn:pmodulus3}
\tilde{\phi_i} = S_i = 0 \ , \\ \nonumber
X_{ji} = 0 \ , \\ \nonumber
\text{ and } \phi_i \text{ arbitrary. } 
\end{eqnarray}
There also exists a runaway direction along which SUSY is restored.
Up to global symmetry transformations it is given by:
\begin{eqnarray}
\label{eqn:runaway3}
\phi_1 = -\left( \frac{mn^2 S_1^2}{\lambda^2 \mu^2} \right)^{1/3},~~~~~
X_{11} = \left( \frac{\mu^2 n S_1}{\lambda m} \right)^{1/3},~~~~~
\tilde{\phi}_1 = \left( \frac{\mu^4 m}{\lambda^2 n S_1} \right)^{1/3}, 
\end{eqnarray}
with $\phi_i$, $\tilde{\phi}_i$, and $X_{ii}$ with $i \neq 1$ given
by~\eqnref{pmodulus3}.  Using global symmetry transformations, we can
rotate any $\phi_i$ vev into $\phi_1$ and $\phi_2$.  Assuming that $\langle
\phi_2 \rangle = 0$, the superpotential for $\phi_1$ is similar to the
R-symmetry breaking model discussed in~\cite{Shih:2007av}, where it
was shown that a metastable minimum may exist near the origin of the
moduli space if (in our notation)
\begin{equation}
\begin{split}
\label{eqn:conditions}
\frac{\lambda^2\mu^4}{m^2 n^2} &< 1  \\  
\frac{\lambda^2\mu^4 - m^2 n^2}{2 m \lambda^2 \mu^2} 
< \langle \phi_1 \rangle &< \frac{m^2 n^2
  - \lambda^2 \mu^4}{2 m \lambda^2 \mu^2} \ .
\end{split}
\end{equation}
In terms of the parameters of the microscopic theory, these relations
imply that $c_B\ll 1$.  We now calculate the one loop correction to
the potential to determine the mass of the pseudomoduli fields.  A
numerical calculation with arbitrary $\phi_1$ and $\phi_2$ shows a
minimum of the Coleman-Weinberg potential at $\phi_2 = 0$.  Thus, we
expand the potential around the $\phi_2 = 0$ in order to obtain an
analytic expression for the mass of $\phi_1$.  We find, in agreement
with~\cite{Shih:2007av},
\begin{equation}
m_{\phi_1}^2 = -\frac{\lambda^2\mu^4}{8\pi^2}\frac{-3m^4 + 2m^2n^2 + n^4 
+ 2m^2(m^2 + 3n^2) \log (\frac{m}{n})}{(m^2-n^2)^3} 
+ N f (\mu^2, \lambda, m, n),
\end{equation}
where $f(\mu^2, \lambda, m, n)$ is strictly positive.  For an
appropriate choice of parameters, the pseudomodulus obtains a non-zero
vacuum expectation value thus breaking the R-symmetry.
In~\figref{modelBCW}, we show the Coleman-Weinberg potential for a few
different parameter choices, demonstrating that for a suitable choice
of input UV parameters, we can drive the pseudomodulus $\phi_1$ to
attain nonzero vev and break $R$-symmetry.

We now turn our attention to the $\det M$ term present in the full
dynamical superpotential. We note that it does not destabilize the
location of the SUSY breaking vacuum since it vanishes in the vicinity
of this minimum. Interestingly, $\det M$ is also vanishing along the
runaway direction. This seems to imply that the runaway behavior
persists in the full theory. On the other hand, the analysis of the
tree-level superpotential shows that the only classical flat
directions are associated with gauge singlets $S_i$. Thus there can be
no runaways with large vevs for composites $M$, $B$, and $\bar
B$. This apparent contradiction is resolved by a careful examination
of the vevs in~\eqnref{runaway3} to $\Lambda$ and the UV cutoff
$\Luv$. In terms of composites of the microscopic theory, we find
\begin{equation}
\begin{split}
\frac{B_1}{\Lambda^N} \sim \left( \frac{\Lambda}{\Luv}
\right)^{\frac{N-2}{3}} \left( \frac{S}{\Luv}\right)^\frac{2}{3} \lsim
\left( \frac{\Lambda}{\Luv} \right)^{\frac{N-2}{3}} \\ 
\frac{M}{\Lambda^2}\sim \left(\frac{\Lambda}{\Luv} 
\right)^{\frac{2N-7}{3}} \left( \frac{S}{\Luv}\right)^\frac{1}{3} \lsim 
\left( \frac{\Lambda}{\Luv} \right)^{\frac{2N-7}{3}} \ , \\
\end{split}
\end{equation}
where we chose the maximal value for $c_B$ and in the second
inequality on each line we used the fact that even in terms of
elementary quarks our theory is only an effective description valid
below $\Luv$. We conclude that along the runaway direction this
effective theory breaks down before the perturbative regime of the
$SU(N)$ gauge dynamics is reached. Thus the reliable determination of
the global supersymmetric minimum of the model requires one to specify
the origin of the non-renormalizable terms in~\eqnref{uvB}.

\subsection{An anomalous R-symmetry in a deformation of an ISS model}
\label{subsec:modelC}

It is instructive to consider another effective model given by the
superpotential
\begin{equation}
W = \lambda \phi_i X^{ij} \tilde{\phi}_j - \mu^2 \text{ Tr }X +
\frac{1}{2} m \phi_i^2 + n \tilde{\phi}_i S^i \ ,
\label{eqn:modelC}
\end{equation}
where $i,j = 1, \ldots F$ are flavor indices as before.  The
R-symmetry charges of the chiral superfields are given by~\footnote{As
  usual, the choice of R-charges is not unique for some fields.}
\begin{equation}
R_X = 2 \ , \quad R_\phi = 1 \ , \quad R_{\tilde{\phi}} = -1 \ , \quad
R_S = 3 \ .
\end{equation}

It is easy to see that once again the model possesses both runaway and
pseudoflat directions in the field space. The pseudoflat direction is
parametrized by
\begin{equation}
\begin{split}
\label{eqn:modelCflat}
\phi_i = \tilde{\phi}_i = S_i = 0  \\
X_{ij} \text{ arbitrary} \ ,
\end{split}
\end{equation}
and the energy along this direction is $V_{min} = (N+1) \mu^4$, where
$F = N+1$.  Most of the $X$ fields are perturbatively stabilized at
the origin. On the other hand, the Coleman-Weinberg potential for
$X_{11}$ is the same as the potential for the pseudomodulus in the
model of~\cite{Shih:2007av}. In particular for $r\equiv m/n\gtrsim
2.11$, the mass of $X_{11}$ is negative and a local minimum with
spontaneously broken R-symmetry exists.

The runaway direction is given by
\begin{equation}
\label{eqn:modelCrunaway}
X_{11} =- \left( \frac{m n^2 S_1^2}{\lambda^2 \mu^2} \right)^{1/3}\,,~~~~~  
\phi_1 = \left( \frac{\mu^2 n S_1}{\lambda m} \right)^{1/3}\,, ~~~~~
\tilde{\phi}_1 = \left( \frac{\mu^4 m}{\lambda^2 n S_1} \right)^{1/3} \ , 
\end{equation}
and in the limit $S_1 \rightarrow \infty$, the vacuum energy is
lowered to $V_{min} = N \mu^4$. Once again, the remaining $X$ fields
are stabilized at the origin by perturbative corrections.

\begin{figure}[htb]
\caption{[color online]. Relative Coleman-Weinberg potential as a
  function of pseudomodulus Tr $X$ for the model in~\eqnref{uv2} where
  $\Lambda = 1$, $\Lambda_{UV} = 10$, $N = 2$, $\lambda = 1$, $m_Q =
  0.1$, $d_B = 8$, and $d_{\bar{B}}$ varied from 0.50 (green, dashed, bottom)
  to 0.70 (purple, solid, middle) to 0.90 (blue, dotted, top).}
\label{fig:modelCCW}
\begin{center}
\includegraphics[width = 0.5\textwidth]{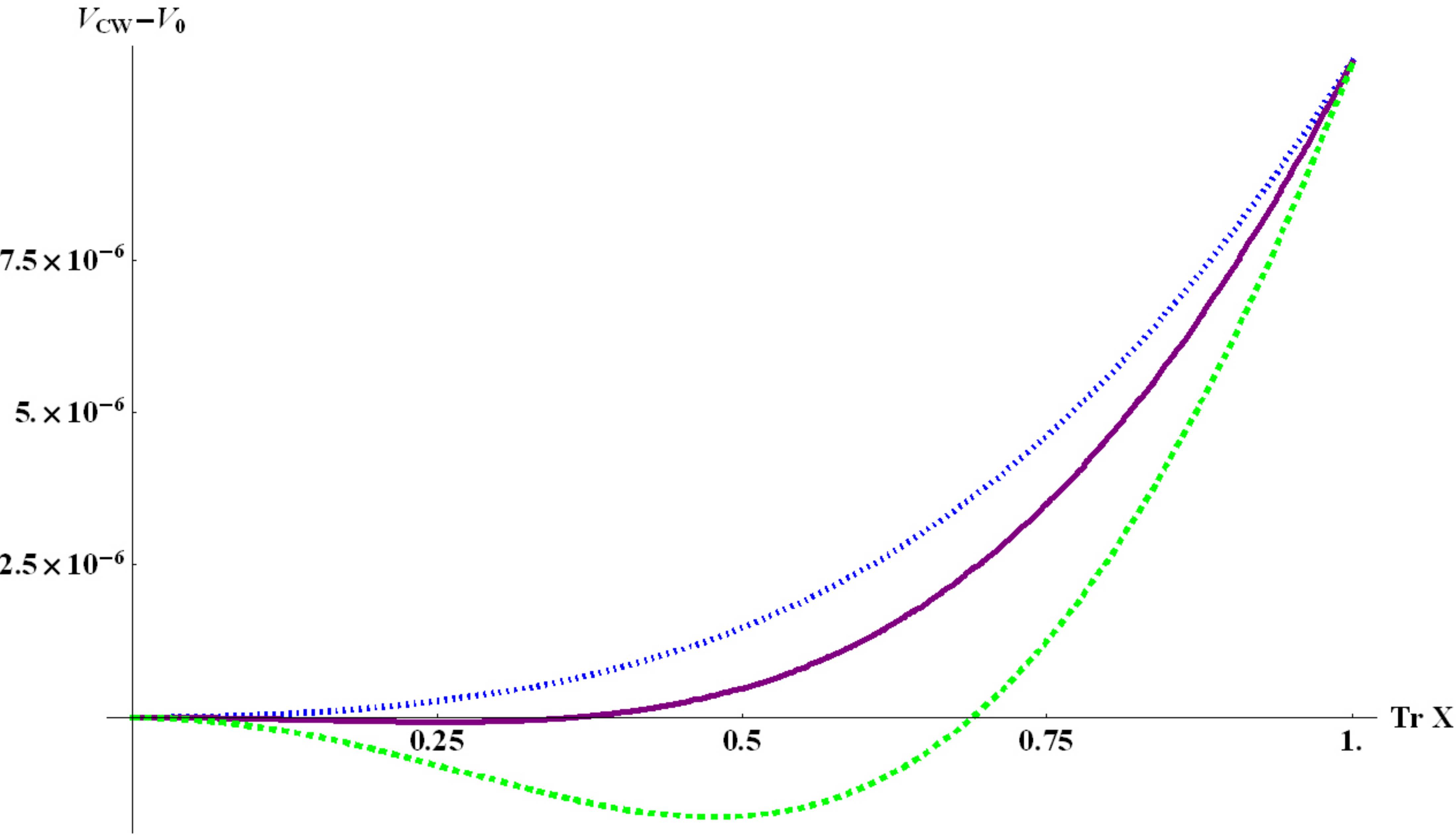}
\end{center}
\end{figure}

We now turn to the analysis of the UV completion of this model.  Once
again we will look for a microscopic description in terms of a
deformation of an s-confining SQCD. The association between gauge
invariant composites of the microscopic description and the fields of
the low energy model is the same as before, see~\eqnref{fieldmap}.
The full non-perturbative superpotential of the UV complete
description is given by
\begin{equation}
\label{eqn:uv2}
W = \lambda \frac{B_i M^{ij} \bar{B}_j - \det M}{\Lambda^{2N-1}} 
 + m_Q \text{ Tr } M + d_B \frac{B_i^2}{\Luv^{2N-3}} 
 + d_{\bar{B}} \frac{\bar{B}_i S^i}{\Luv^{N-2}} \ .
\end{equation}
In this case, the parameters of the UV and IR descriptions are related
by
\begin{equation}
\mu^2 = m_Q \Lambda \ , \quad \frac{m}{2} = d_B
\left(\frac{\Lambda}{\Luv} \right)^{2N-3} \Lambda \ , \quad n =
d_{\bar{B}} \left( \frac{\Lambda}{\Luv} \right)^{N-2} \Lambda \ ,
\end{equation}
and $\lambda \sim \mathcal{O}(1)$.  Requiring that R-symmetry is
spontaneously broken we find
\begin{equation}
 r = 2
\frac{d_B}{d_{\bar{B}}} \left( \frac{\Lambda}{\Lambda_{UV}}
\right)^{N-1} \gtrsim 2.11\,.
\end{equation}
Thus R-symmetry breaking requires a mild hierarchy between parameters
of tree level superpotential $d_B>d_{\bar{B}}$.  The Coleman-Weinberg
potential for several choices of the parameters is shown
in~\figref{modelCCW}.

The analysis of the symmetries in this model is analogous to that in
Sec.~\ref{subsec:modelB}.  Tree level terms in the superpotential are
invariant under an anomalous $U(1)_R$ symmetry given by
\begin{equation}
U(1)^\prime_{R} = U(1)_R + N U(1)_A - (N - 1) U(1)_B \ .
\end{equation}
Furthermore, assigning the charge $2N$ to the dynamical scale
$\Lambda^{2N-1}$ makes the full non-perturbative potential invariant
under this symmetry.  The R-symmetry matching between the UV and IR is
again given by ~\eqnref{Rmatch}. Once again, there are no fields with
negative R-charges in the microscopic description.

We conclude this section by discussing the restoration of SUSY in the
microscopic description.  Recall that there are no supersymmetric
ground states in the low energy description given
by~\eqnref{modelC}. Moreover, $\det M\sim \det X$ vanishes along the
non-supersymmetric runaway~\eqnref{modelCrunaway}. However, the
presence of this non-perturbatively generated term leads to the
appearance of supersymmetric vacua elsewhere.

\section{Conclusions}
\oraf-like models with spontaneously broken R-symmetry require that
some of the IR degrees of freedom carry negative R-charges. Therefore
symmetries of the low energy description allow superpotential terms
that are singular at the origin of the moduli space. Such terms could,
in principle, be generated by non-perturbative dynamics of the UV
theory and, if present, would be dangerous.  This possibility
underscores the importance of finding UV completions of
phenomenologically viable models.  In this paper we have considered
several generalizations of models introduced in~\cite{Shih:2007av} and
constructed their UV completions. We have shown that an R-symmetry of
the effective low energy description can be mapped either to an
anomaly-free or anomalous R-symmetry of the microscopic physics.  In
the former case, the R-symmetry of the IR description is a linear
combination of R and non-R symmetries of the UV physics.  In the
latter case, the negative R-charges in the IR description are due to
the anomaly --- specifically, the contribution of the spurion ---
while all the elementary fields carry non-negative R-charge.  In
either case, the existence of the anomaly-free non-R symmetry forbids
the appearance of dangerous terms in the dynamical superpotential.  It
is interesting to note that some models of direct gauge mediation
(see, for example,~\cite{Csaki:2006wi}) possess an anomalous
R-symmetry which is broken perturbatively through the mechanism
of~\cite{Shih:2007av}.

We have shown that in successful UV completions the dynamics of the
model in the vicinity of the SUSY breaking ground state usually can be
analyzed reliably in terms of the low energy description. On the other
hand, the location (or even existence) of the supersymmetric ground
state depends sensitively on the details of the microscopic
physics. Thus several important issues, such as the lifetime of the
SUSY breaking vacuum and the cosmological history of the model, cannot
be reliably analyzed within the low energy approximation.

\section{Acknowledgements}
\label{sec:acknowledgements}
The work of J.G., M.I., Y.S., and F.Y. was supported in part by the
National Science Foundation under the grants No. PHY-0653656 and
PHY-0970173.  F.Y. was also supported by a 2010 LHC Theory Initiative
Graduate Fellowship, NSF Grant No. PHY-0705682.

 
\end{document}